\documentclass[twocolumn,floatfix,aps,prb]{revtex4}
\usepackage{graphicx}
\usepackage{amsmath}
\usepackage{amssymb}
\usepackage{bm}
\usepackage{hyperref}

\usepackage{color}

\begin{document}

\title{Topologically protected charge transfer along the edge of a chiral \textit{p}-wave superconductor}
\author{N. V. Gnezdilov}
\affiliation{Instituut-Lorentz, Universiteit Leiden, P.O. Box 9506, 2300 RA Leiden, The Netherlands}
\author{B. van Heck}
\affiliation{Instituut-Lorentz, Universiteit Leiden, P.O. Box 9506, 2300 RA Leiden, The Netherlands}
\author{M. Diez}
\affiliation{Instituut-Lorentz, Universiteit Leiden, P.O. Box 9506, 2300 RA Leiden, The Netherlands}
\author{Jimmy A. Hutasoit}
\affiliation{Instituut-Lorentz, Universiteit Leiden, P.O. Box 9506, 2300 RA Leiden, The Netherlands}
\author{C. W. J. Beenakker}
\affiliation{Instituut-Lorentz, Universiteit Leiden, P.O. Box 9506, 2300 RA Leiden, The Netherlands}
\date{September 2015}
\begin{abstract}
The Majorana fermions propagating along the edge of a topological superconductor with $p_x+ip_y$ pairing deliver a shot noise power of $\frac{1}{2}\times e^2/h$ per eV of voltage bias. We calculate the full counting statistics of the transferred charge and find that it becomes trinomial in the low-temperature limit, distinct from the binomial statistics of charge-$e$ transfer in a single-mode nanowire or charge-$2e$ transfer through a normal-superconductor interface. All even-order correlators of current fluctuations have a universal quantized value, insensitive to disorder and decoherence. These electrical signatures are experimentally accessible, because they persist for temperatures and voltages large compared to the Thouless energy.
\end{abstract}
\maketitle

The chiral edge modes of the quantum Hall effect have a superconducting analogue in chiral \textit{p}-wave superconductors, with a spin-triplet $p_x+ip_y$ pair potential.\cite{Sen00,Rea00} A temperature gradient is predicted to drive a heat current along the edge carried by Majorana fermions, equal-weight superpositions of electron and hole excitations in the superconducting gap. The thermal conductance $G$ is quantized at the electronic quantum $G_0=\pi^2 k_{\rm B}^2T/3h$ times one-half, loosely speaking because one electron equals to two Majorana fermions --- or more fundamentally\cite{Rea00,Sum13,Bra15} because the field theory of Majorana edge modes has topological (central) charge $c=1/2$. 

Quantized electrical signatures of the Majorana edge mode are lacking, basically because Majorana fermions are charge neutral. It has been argued\cite{Hua14} that the lack of topological protection of charge currents is the reason that experiments\cite{Hic10,Jan11,Cur14} on ${\rm Sr}_2{\rm RuO}_4$ have not found the predicted magnetic moment of a circulating edge current.\cite{Mat99} A domain wall between opposite chiralities $p_x\pm ip_y$ of the pair potential has a nonzero electrical conductance, but its value is not quantized.\cite{Ser10}

\begin{figure}[tb]
\centerline{\includegraphics[width=0.9\linewidth]{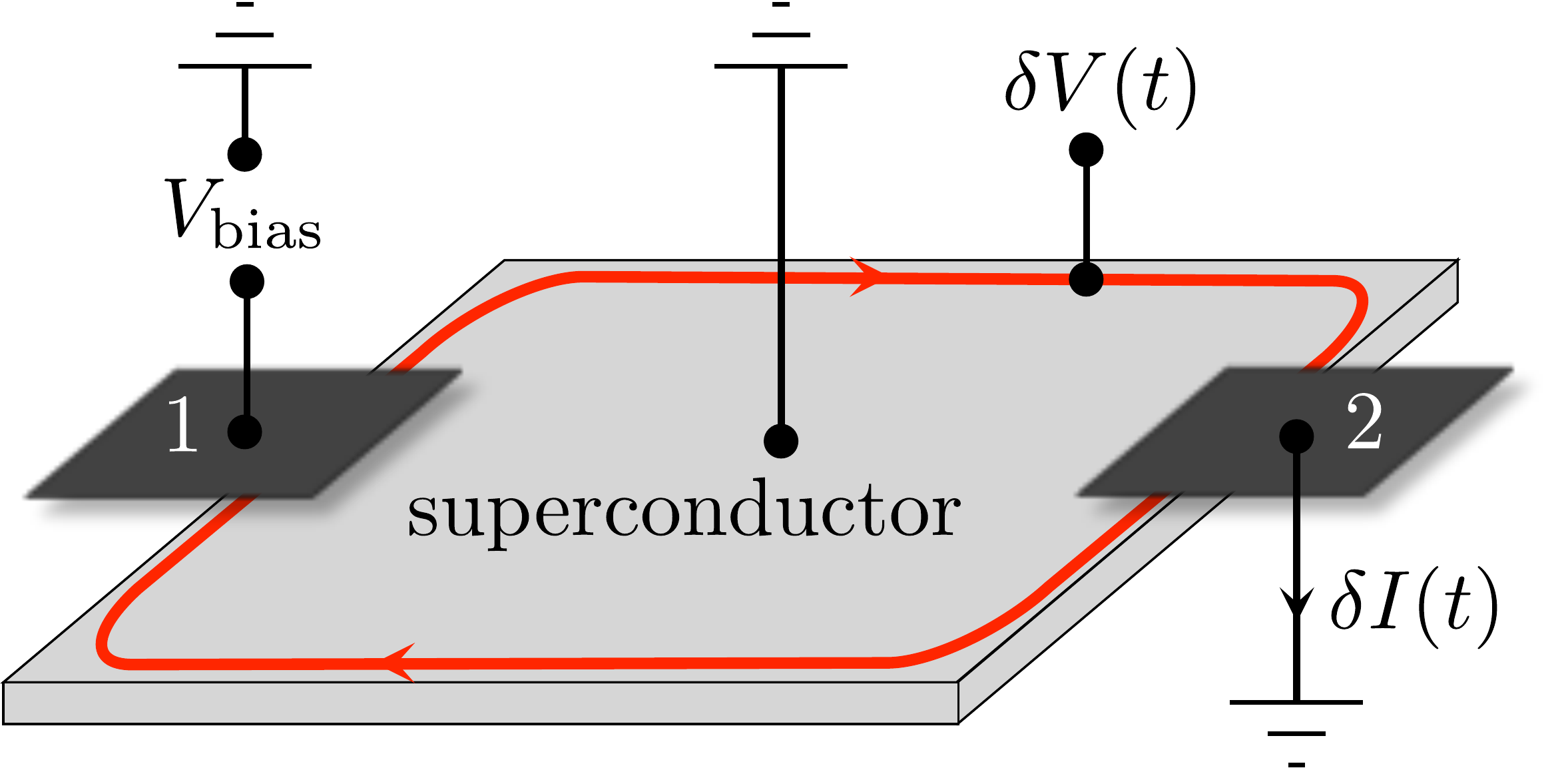}}
\caption{Nonlocal current and voltage measurement to detect the charge-neutral Majorana edge mode in a two-dimensional topological superconductor. A bias voltage $V$ excites the edge mode, producing a fluctuating current $\delta I(t)$ and voltage $\delta V(t)$, detected at a remote contact. Because the bulk of the superconductor is grounded, these nonlocal signals are evidence for conduction by gapless edge excitations.
}
\label{fig_layout}
\end{figure}

Although the electrical conductance of a Majorana mode vanishes, the electrical shot noise power is nonzero.\cite{Akh11,Die14} Particle-hole-symmetry enforces a one-to-one relationship between the zero-temperature shot noise power $P$ and the thermal conductance,
\begin{equation}
P/P_0=G/G_0=\tfrac{1}{2}\,{\rm Tr}\,tt^\dagger,\;\;P_0=e^3V/h,\label{PGrelation}
\end{equation}
where $t$ is the rank-one transmission matrix between two contacts along the edge (biased at voltage $V>0$, see Fig.\ \ref{fig_layout}). By definition,\cite{Bla99,Bee03,note0}
\begin{equation}
P=\int_{-\infty}^\infty dt\,\langle\delta I(0)\delta I(t)\rangle=\tau^{-1}\,{\rm Var}\,q,\label{Pdef}
\end{equation}
the shot noise power is the correlator of the current fluctuations and gives the variance of the charge transferred between the contacts in a time $\tau$. Eq.\ \eqref{PGrelation} implies that ${\rm Var}\,q$ has the universal value $\frac{1}{2}P_0 \tau$ for a fully transmitted Majorana mode.\cite{Akh11}

Eq.\ \eqref{PGrelation} says that the second moment of the transferred charge in a Majorana mode is directly determined by the quantized thermal conductance. Higher moments are not so constrained, and one might ask whether they are quantized as well. Here we calculate the full probability distribution $P(q)$ of the transferred charge, including also the effects of finite temperature. In the zero-temperature limit we find that the characteristic (moment generating) function $F(\chi)=\langle e^{i\chi q}\rangle$, related to $P(q)$ by a Fourier transform, has the form
\begin{equation}
F(\chi)=\bigl[1+\tfrac{1}{4}(e^{ie\chi}-1)+\tfrac{1}{4}(e^{-ie\chi}-1)\bigr]^{\cal N}.\label{Fchitrinomial}
\end{equation}
This describes a trinomial counting statistics where ${\cal N}=eV\tau/h$ attempts transfer either $-e$, $0$, or $+e$ charge, with probabilities $1/4$, $1/2$, and $1/4$, respectively. 

This result for the statistics of charge transported by a chiral Majorana edge mode can be contrasted with the characteristic function for the charge transmitted by an electronic mode in a nanowire,\cite{Lev93}
\begin{equation}
F_{\rm electron}(\chi)=\bigl[1+{\cal T}(e^{ie\chi}-1)\bigr]^{\cal N},\label{FchiLL}
\end{equation}
where ${\cal T}\in[0,1]$ is the transmission probability. The two key distinctions with Eq.\ \eqref{Fchitrinomial} are that the counting statistics is binomial, rather than trinomial, and that the transfer probability is not quantized. A chiral quantum Hall edge mode would have a quantized ${\cal T}=1$, but then there would be no charge fluctuations at all. 

To see that these distinctions are not merely a consequence of the presence of a superconductor, we compare with the corresponding result for the charge transferred through a normal-metal-superconducting (NS) point contact,\cite{Muz94}
\begin{equation}
F_{\rm NS}(\chi)=\bigl[1+{\cal R}_{\rm A}(e^{2ie\chi}-1)\bigr]^{\cal N}.\label{FchiNS}
\end{equation}
The transmission probability is replaced by the probability ${\cal R}_{\rm A}$ for Andreev reflection and the unit of transferred charge is doubled, but the statistics remains binomial and not quantized.

Resonant tunneling through a Majorana zero-mode, bound to a vortex or to the end of a nanowire, provides another point of comparison.\cite{Li13,Sol14,Liu15} For two contacts biased at voltage $\pm V/2$ and coupled to the zero-mode with tunnel probabilities ${\cal T}_1$, ${\cal T}_2$, the charge entering contact 1 has characteristic function\cite{Liu15}
\begin{equation}
F_{\text{zero-mode}}=\bigl[1+{\cal T}_{1}({\cal T}_{1}+{\cal T}_2)^{-1}(e^{ie\chi}-1)\bigr]^{\cal N}.\label{FchiM}
\end{equation}
The statistics is binomial and dependent on the tunnel probabilities, except for a symmetric junction (when ${\cal T}_1={\cal T}_2$ it drops out).\cite{note1}

Our analysis follows the scattering theory of counting statistics pioneered by Levitov and Lesovik,\cite{Lev93} in the convenient formulation of Klich.\cite{Kli02} The characteristic function is given by
\begin{align}
F(\chi) ={}& {\rm Tr}\,\rho_0 \exp \left[ ie\chi \sum_{E>0} c^\dagger(E){\cal P}c(E) \right] \nonumber\\
&\times\exp \left[ -ie\chi \sum_{E>0} c^\dagger(E){\cal M}(E)c(E) \right] , \label{chiT} \\
{\cal M}(E)={}&S^\dagger(E){\cal P}S(E),\;\;S=\begin{pmatrix}
r'&t'\\
t&r
\end{pmatrix},\;\;{\cal P}=\begin{pmatrix}
0&0\\
0&\sigma_z
\end{pmatrix}.\label{MPdef}
\end{align}
The sum over energies is understood as $\sum_{E}\rightarrow(\tau/h)\int dE$ in the limit $\tau\rightarrow\infty$. The trace gives the expectation value with density matrix $\rho_0$ of the fermion operators $c,c^\dagger$, representing the quasiparticles injected into the edge from the two contacts, contact 1 at voltage $V$ and contact 2 grounded. The scattering matrix $S$ relates incoming and outgoing quasiparticles, with reflection and transmission subblocks. The matrix ${\cal P}$ selects the quasiparticles at contact 2, where the current is measured. The Pauli matrix $\sigma_z$  appearing in ${\cal P}$ acts on the electron-hole degree of freedom, to account for the fact that electron and hole quasiparticles contribute with opposite sign to the electrical current.

Because different energies are uncoupled, we may perform the trace at each energy separately, so that we may write Eq.\ \eqref{chiT} in the form
\begin{align}
&\ln F(\chi) = \sum_{E>0}\ln {\rm Tr}\biggl(e^{-\beta c^{\dagger}(E){\cal E}(E)c(E)}  \nonumber\\
&\quad\times e^{ ie\chi c^\dagger(E){\cal P}c(E)}e^{ -ie\chi c^\dagger(E){\cal M}(E)c(E)}\biggr)-\ln Z, \label{chiT2}\\
&{\cal E}(E)=E-eV\begin{pmatrix}
\sigma_z&0\\
0&0
\end{pmatrix}.\label{calEdef}
\end{align}
Here $\beta=1/k_{\rm B}T$ and $Z={\rm Tr}\,e^{-\beta c^\dagger{\cal E}c}$ is the partition function at temperature $T$ (the same for both contacts) and chemical potential $\mu$ (equal to $\pm eV$ for electrons and holes at contact 1, and equal to 0 at contact 2). 

With the help of the formula\cite{Kli02,Avr08}
\begin{equation}
{\rm Tr}\,\textstyle{\prod_n} e^{c^\dagger A_nc}={\rm Det}\,\left(1+\textstyle{\prod_n} e^{A_n}\right),\label{Klich}
\end{equation}
the expression \eqref{chiT2} reduces to
\begin{align}
&\ln F(\chi) = \frac{\tau}{h}\int_{0}^\infty dE\,\ln {\rm Det}\bigl(1-{\cal F}+{\cal F}e^{ ie\chi {\cal P}}e^{ -ie\chi {\cal M}}\bigr), \label{chiT3}\\
&{\cal F}(E)=[1+e^{\beta{\cal E}(E)}]^{-1}.\label{calFdef}
\end{align}
The matrix exponentials simplify because ${\cal P}^{2n}={\cal P}^2$, ${\cal P}^{2n+1}={\cal P}$, ${\cal M}^{2n}={\cal M}^2$, ${\cal M}^{2n+1}={\cal M}$ (in view of unitarity, $SS^{\dagger}=1$), hence
\begin{equation}
\begin{split}
&e^{ie\chi{\cal P}}=1+{\cal P}^2(\cos e\chi-1)+i{\cal P}\sin e\chi,\\
&e^{-ie\chi{\cal M}}=1+{\cal M}^2(\cos e\chi-1)-i{\cal M}\sin e\chi.
\end{split}\label{simplifyexp}
\end{equation}
In the zero-temperature limit ${\cal F}(E){\cal P}\rightarrow 0$ for $E>0$, so the factor $e^{ie\chi{\cal P}}$ in Eq.\ \eqref{chiT3} may be replaced by unity. To first order in $V$ we then have
\begin{align}
&\ln F_0(\chi) = (eV\tau/h)\ln {\rm Det}\bigl[1+\nonumber\\
&\qquad\qquad\mbox{}+\tfrac{1}{2}(1+\sigma_z) t^\dagger\bigl(\cos e\chi-1-i\sigma_z\sin e\chi\bigr) t\bigr],\label{F0chi}
\end{align}
with transmission matrix $t$ evaluated at the Fermi energy $E=0$.

These formulas hold for any channel connecting two metal contacts via a superconductor. We now use that the connection is via an unpaired Majorana edge mode, which implies that the $2N\times 2N$ transmission matrix has rank one, irrespective of the number $2N$ of electron-hole modes in the metal contact: $t={\cal T}^{1/2}\,uv^{\rm T}$ with unit vectors $u,v$ and transmission probability ${\cal T}$. Particle-hole symmetry at the Fermi level requires that $t=\sigma_x t^\ast\sigma_x$, hence the matrix $t^\dagger \sigma_z t$ vanishes identically:
\begin{equation}
t^\dagger \sigma_z t=-i\sigma_x t^{\rm T}\sigma_y t=-i{\cal T}(u^{\rm T}\sigma_y u)\sigma_x vv^{\rm T}=0.\label{tdaggersigmazt}
\end{equation}
Similarly, $t\sigma_z t^\dagger=0$ while $tt^\dagger$ has a single nonzero eigenvalue equal to ${\cal T}$. We thus arrive at
\begin{equation}
\ln F_0(\chi) = {\cal N}\ln \bigl[1+\tfrac{1}{2}{\cal T}(\cos e\chi-1)],\;\;{\cal N}=eV\tau/h,\label{F0chiresult}
\end{equation}
which for ${\cal T}=1$ is the result \eqref{Fchitrinomial} announced in the introduction. 

The corresponding trinomial probability distribution $P(q)$ of the transferred charge $q=0,\pm 1,\pm 2,\ldots \pm {\cal N}$ (in units of the electron charge $e$) is given by
\begin{align}
P(q)={}&\frac{(2-{\cal T})^{{\cal N}-|q|}{\cal T}^{|q|}{\cal N}!}{2^{{\cal N}+|q|}({\cal N}-|q|)!|q|!}\nonumber\\
&\times\mbox{} _{2}F_1\left[\textstyle{\frac{|q|-{\cal N}}{2},\frac{|q|+1-{\cal N}}{2},|q|+1,\frac{{\cal T}^2}{(2-{\cal T})^2}}\right],\label{Pqgeneral}
\end{align}
with $_{2}F_1$ the hypergeometric function. For ${\cal T}=1$ this simplifies to
\begin{equation}
P(q)=2^{-2{\cal N}}{{2{\cal N}}\choose{{\cal N}-q}},\label{Pqresult}
\end{equation}
which looks like a \textit{displaced} binomial distribution.\cite{note1} Cumulants are coefficients in the Taylor series $\ln F(\chi)=\sum_{p} \langle\!\langle q^p\rangle\!\rangle(i\chi)^p/p!$, giving for ${\cal T}=1$ the result
\begin{equation}
\langle\!\langle q^p\rangle\!\rangle=\begin{cases}
2{\cal N}(2^{p+1}-1)\frac{p!}{(p+1)!}B_{p+1}&\mbox{for $p$ even},\\
0&\mbox{for $p$ odd},
\end{cases}\label{cumulant}
\end{equation}
with $B_{p+1}$ the Bernoulli number. The first few values are
\begin{equation}
{\cal N}^{-1}\langle\!\langle q^p\rangle\!\rangle=\tfrac{1}{2},-\tfrac{1}{4},\tfrac{1}{2},-\tfrac{17}{8},\tfrac{31}{2}\;\;{\rm  for}\;\; p=2,4,6,8,10.\label{Bernoulli}
\end{equation}

\begin{figure}[tb]
\centerline{\includegraphics[width=0.7\linewidth]{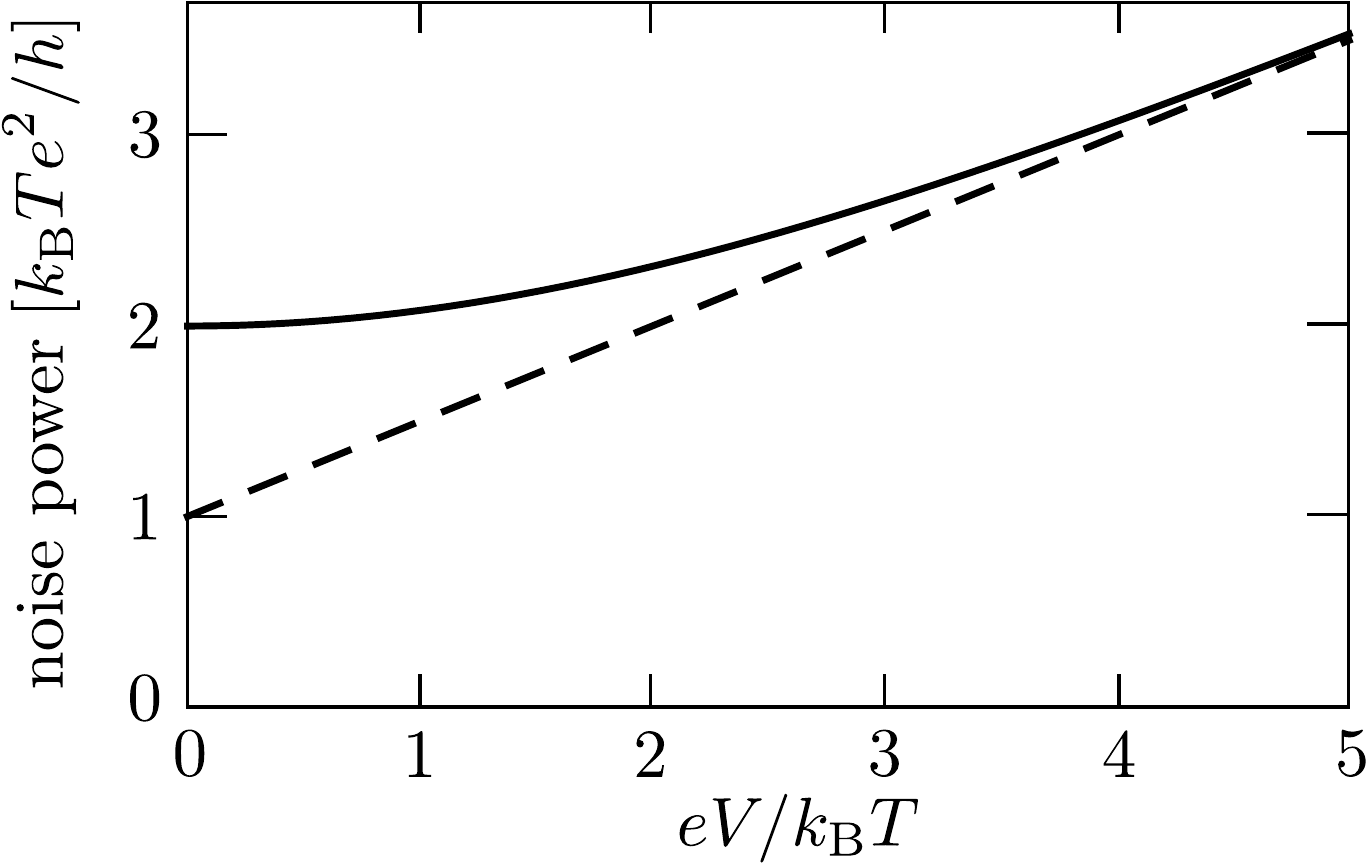}}
\caption{Noise power $P$ of the Majorana edge mode at temperature $T$, as a function of bias voltage $V$. The solid curve is calculated from Eq.\ \eqref{Pthermalshotexact}, the dashed line is the low-temperature asymptote \eqref{Pthermalshot}. These are results for a single-channel contact ($N=1$) to a fully transmitted edge mode (${\cal T}=1$). Here the voltage is assumed to be small on the scale of the superconducting gap $\Delta_0$. See Fig.\ \ref{fig_numerics} for the voltage dependence in the regime $k_{\rm B}T\ll eV\lesssim\Delta_0$.
}
\label{fig_noise}
\end{figure}

All of this is at zero-temperature. The general finite-temperature formulas are complicated, for a compact expression we take the case $N=1$, ${\cal T}=1$ of a single-channel contact to a fully transmitted edge mode. The determinant in Eq.\ \eqref{chiT3} then evaluates to
\begin{align}
&\ln F(\chi) = \frac{\tau}{h}\int_{0}^\infty dE\,\ln \bigl[1+\tfrac{1}{2}f(1-f)(\cos 2e\chi-1)\nonumber\\
&\qquad\qquad+(f+\tfrac{1}{2}f_V-ff_V)(\cos e\chi-1)\bigr],\label{Fchifinite}\\
&f(E)=(1+e^{\beta E})^{-1},\;\;f_V=f(E-eV)+f(E+eV).\label{f12def}
\end{align}
This is the multinomial statistics of transferred charge $0,\pm e,\pm 2e$, with the interpretation that charge $\pm e$ is transferred via the Majorana edge mode and charge $\pm 2e$ is transferred via Andreev reflection into the bulk superconductor. The corresponding noise power is
\begin{align}
P&=-\frac{1}{\tau}\lim_{\chi\rightarrow 0}\frac{d^2}{d\chi^2}\ln F(\chi)\nonumber\\
&=\frac{e^2}{h}\int_0^\infty dE\,\bigl[\tfrac{1}{2}f_V(1-2f)+3f-2f^2\bigr]\label{Pthermalshotexact}\\
&=\frac{e^2}{h}\times\begin{cases}
(k_{\rm B}T+\tfrac{1}{2}eV)&{\rm for}\;\;k_{\rm B}T\lesssim eV,\\
2k_{\rm B}T&{\rm for}\;\;eV\ll k_{\rm B}T.
\end{cases}\label{Pthermalshot}
\end{align}
As a check, we note that the thermal noise power is related to the contact conductance\cite{note2} $G=e^2/h$ by
\begin{equation}
P_{\rm thermal}\equiv\lim_{V\rightarrow 0}P=2k_{\rm B}TG, \label{Nyquis}
\end{equation}
in accordance with the Johnson-Nyquist relation.\cite{note0} 

\begin{figure}[tb]
\centerline{\includegraphics[width=0.9\linewidth]{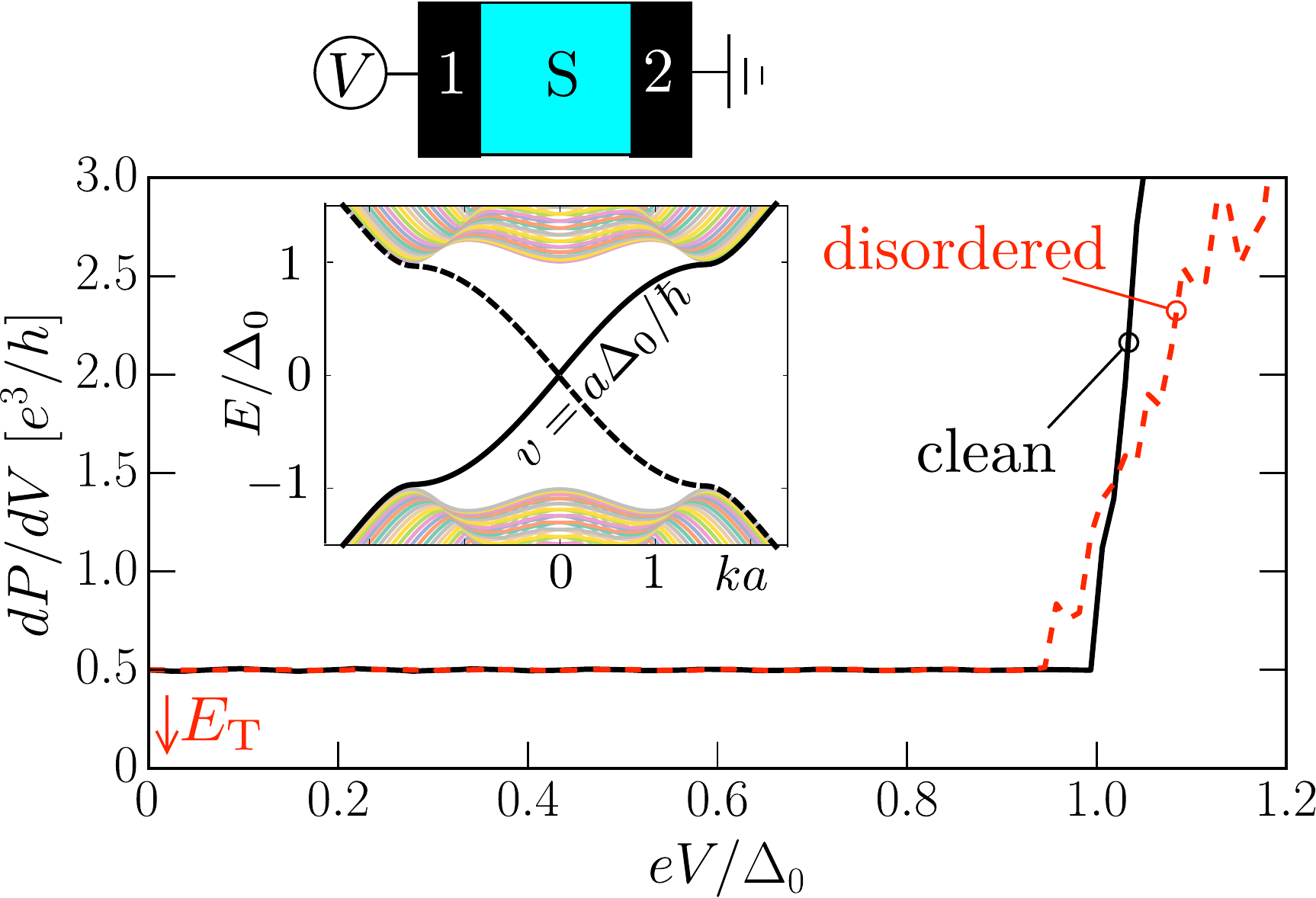}}
\caption{Voltage dependence of the zero-temperature shot noise power, calculated numerically \cite{KWANT} for the tight-binding Hamiltonian of a chiral \textit{p}-wave superconductor (black solid and dashed curves in the band structure are the counterpropagating Majorana modes at opposite edges). The superconducting region ($W=L=100$ in units of the lattice constant $a$) is connected to metallic leads (superconductor and lead 2 grounded, lead 1 at voltage $V$). The electron-to-electron and electron-to-hole transmission matrices $t_{ee}(E)$ and $t_{he}(E)$ from lead 1 to 2 are calculated as a function of the energy $E=eV$. The differential shot noise power in lead 2 then follows from\cite{Ana96} $dP/dV=(e^3/h){\rm Tr}\,(T_+-T_-^2)$, with $T_\pm=t_{ee}^\dagger t_{ee}\pm t_{he}^\dagger t_{he}$. The black solid curve is the result for a clean system, the red dashed curve is obtained in the presence of a random on-site potential $U_n\in[-\Delta_0/2,\Delta_0/2]$. The red arrow indicates the Thouless energy $E_{\rm T}=\hbar v/L$. This simulation demonstrates that the quantized shot noise $dP/dV=e^3/2h$ is insensitive to disorder for voltages $|V|\lesssim\Delta_0/e$ --- even if $eV$ is large compared to $E_{\rm T}$.
}
\label{fig_numerics}
\end{figure}

From Fig.\ \ref{fig_noise} we see that the slope $dP/dV$ is within $10\%$ of the quantized value $e^2/2h$ for voltages $eV\gtrsim 3k_{\rm B}T$. This lower limit on the voltage is to be combined with an upper limit set by the superconducting energy gap. From Fig.\ \ref{fig_numerics} we estimate that the quantization is preserved even in the presence of strong disorder when $eV\lesssim \Delta_0/2$. For a realistic gap\cite{gap} of $0.2\,{\rm meV}$ the quantized shot-noise regime would then be accessible at temperatures below $0.4\,{\rm K}$, which is a quite feasible requirement.

We discuss three further issues regarding the robustness of the quantized shot noise of the Majorana edge mode. 

1) Impurity scattering along the edge has no effect because of the chirality of the edge mode, prohibiting backscattering. The contact resistance may in principle reduce ${\cal T}$ below unity, but this effect can be minimized by using an extended contact: If each of the two contacts contains $2N$ electron-hole modes with tunnel probability $\Gamma$ to the edge mode, then ${\cal T}\simeq[\min(1,N\Gamma)]^2$. Hence contact resistances have no effect on the quantized shot noise if $N\Gamma\gtrsim 1$. We need to avoid a large thermal noise due to Andreev reflection from such an extended contact, which is of order $k_{\rm B}T N\Gamma^2$. Both conditions, maximal coupling (${\cal T}=1$) with minimal thermal noise ($P_{\rm thermal}\simeq k_{\rm B}Te^2/h$), are satisfied if we take $1/N\ll \Gamma\ll 1/\sqrt{N}$, so that $N\Gamma\gg 1$, $N\Gamma^2\ll 1$. 

2) Loss of phase coherence has no effect on the quantized shot noise. The coherent electron-hole superposition of a Majorana fermion is fragile indeed, coupling to the electromagnetic environment will project it onto an electron or a hole, effectively measuring the charge of the quasiparticle. The trinomial statistics, however, remains unaffected, because each of the ${\cal N}$ current pulses still transfers the same amount of charge $\pm e$ with equal probability $1/4$.

3) The shot noise quantization is a macroscopic effect, preserved on scales large compared to the Thouless energy $E_{\rm T}=\hbar v/L$. This is the energy scale on which electrons and holes dephase after traveling with velocity $v$ over a distance $L$ and which governs transport experiments in an interferometer.\cite{Str11,Str15} As demonstrated in Fig.\ \ref{fig_numerics}, raising the voltage to $eV\approx E_{\rm T}$ has a negligible effect on the shot noise. The reason for this unusual insensitivity is that the electron and hole component of the Majorana mode acquire \textit{the same} phase factor at finite energy,\cite{Sto04} so no dephasing can occur. The fact that the energy scale for the quantization is set by $\Delta_0$ rather than by $E_{\rm T}$ is crucial for the observability of the effect.

In conclusion, we have identified unique electrical signatures of a charge-neutral Majorana mode propagating along the edge of a topological superconductor: A trinomial statistics of transferred charge, with quantized cumulants persisting in a macroscopic system, since they are insensitive to impurity scattering or loss of phase coherence. A promising physical system to search experimentally for the shot noise quantization could be an array of parallel nanowires\cite{Die14,Wan14,Wak14,Ser14,Ful14} or parallel chains of magnetic atoms,\cite{Nad14,Lu15} all on a superconducting substrate.

As a direction for further theoretical research we point to the effect of interactions among the Majorana fermions. Two recent studies \cite{Mil15,Rah15} have found an interaction-driven quantum phase transition from central charge $c=1/2$ to $c=3/2$. Because the coefficient $1/2$ in the shot noise power \eqref{PGrelation} originates from the central charge of the Majorana mode, it would be interesting to see what is the effect of this phase transition on the charge transfer statistics. A related extension of our results would be to topological superconductors with a higher Chern number, supporting multiple Majorana modes at each edge, for which physical realizations have been recently predicted.\cite{Ron14,Sca15}

This research was supported by the Foundation for Fundamental Research on Matter (FOM), the Netherlands Organization for Scientific Research (NWO/OCW), and an ERC Synergy Grant.

\appendix

\section{Effect of a tunnel barrier at the normal-superconductor contact}

\begin{figure}[b]
\centerline{\includegraphics[width=0.9\linewidth]{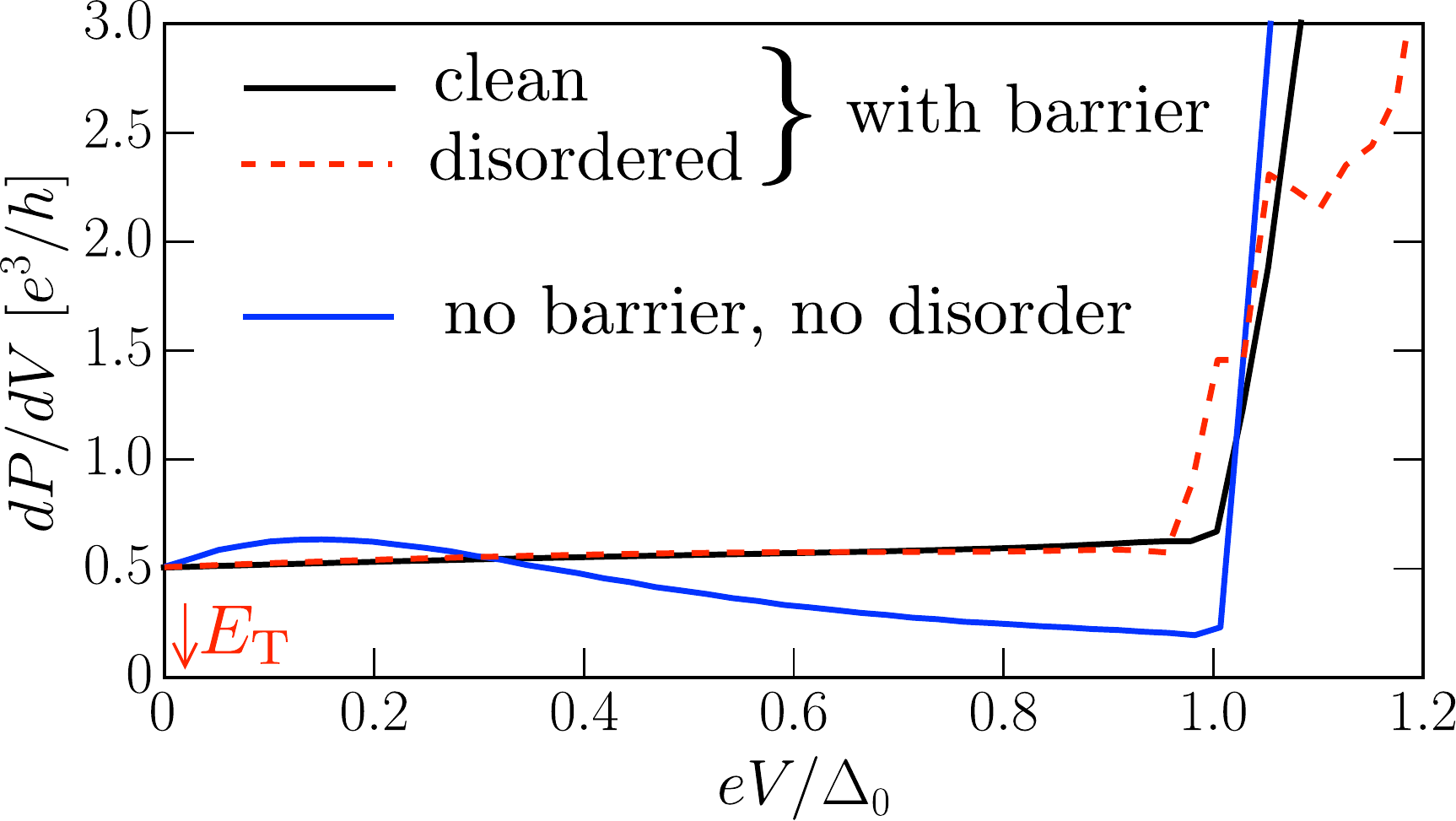}}
\caption{Voltage dependence of the zero-temperature shot noise power in the model of Fig.\ \ref{fig_numerics}. The black curves are with a tunnel barrier at the NS interfaces (transmission probability $\Gamma=0.32$ in each of $N=100$ modes), the blue curve is without any barrier. For $V\lesssim E_{\rm T}$ there is no difference, for larger $V$ a tunnel barrier helps to preserve the quantization all the way up to the superconducting gap. (The black solid curve differs slightly from Fig.\ \ref{fig_numerics}, because of an improvement in the tight-binding modeling of the NS interface.)
}
\label{fig_barrier}
\end{figure}

As we explained in the main text, a tunnel barrier at the normal-metal--superconductor (NS) interface is needed to suppress the thermal noise from Andreev reflection. At zero temperature this is of no concern, but it turns out that a tunnel barrier is still advantageous because it reduces the voltage sensitivity of the shot noise power. We show this in Fig.\ \ref{fig_barrier}, where we compare the zero-temperature $dP/dV$ for the two cases.

\begin{thebibliography}{99}
\bibitem{note0} The dissipated power in a contact resistance $R=1/G$, measured in a frequency band width $\Delta f$, equals $P_{\rm diss}=2RP\Delta f$. To avoid the factor of two, an alternative definition of $P$ has a factor of two in front of the integral in Eq.\ \eqref{Pdef}. With our definition, the Johnson-Nyquist formula for thermal noise is $P_{\rm thermal}=2k_{\rm B}TG$, while $P_{\rm diss}=4k_{\rm B}T\Delta f$.
\bibitem{note1} For a symmetric coupling to the Majorana zero-mode (${\cal T}_1={\cal T}_2$), the trinomial and binomial characteristic functions \eqref{Fchitrinomial} and \eqref{FchiM} are related by $\ln F_{\text{zero-mode}}(\chi)=\frac{1}{2}ie{\cal N}\chi+\frac{1}{2}\ln F(\chi)$, so cumulants of order two and higher differ only by a factor of two. This explains the correspondence between the Bernoulli numbers in Eq.\ \eqref{cumulant} and the Euler polynomial of Ref.\ \onlinecite{Liu15}. For general ${\cal T}_n$ there is no such correspondence between trinomial and binomial distributions.
\bibitem{note2} The contact conductance $G$ in the geometry of Fig.\ \ref{fig_layout} is given in terms of the $2N\times 2N$ reflection matrix $r$ at the contact by $G=\tfrac{1}{2}(e^2/h)\,{\rm Tr}\,(1-r^\dagger\tau_z r\tau_z)$. For $N=1$, ${\cal T}=1$ we have $r=ab^{\rm T}$, $r^\dagger=\sigma_x ba^{\rm T}\sigma_x$, with unit vectors $a,b$, hence $G=e^2/h$.
\end{thebibliography}
\end{document}